\documentstyle[aps,prl,twocolumn]{revtex}
\input epsf

\def \K    {K^+/K^-}
\def \fNy  {\langle f(y) \rangle_{N}^{\rm Kin}}
\def \fN   {\langle f \rangle_{N}^{\rm Kin}}
\def \fBch {\langle f \rangle_{B}^{\rm Ch}}
\def \fBth {\langle f \rangle_{B}^{\rm Kin}}
\def \pn   {{\rm p}/{\rm n}}
\def \npa  {N_{\rm part}}

\begin{document}
\draft


\title{Correlation between the charged kaon ratio and
the baryon phase-space density in heavy-ion collisions}

\author{Fuqiang Wang}
\address{Nuclear Science Division, 
	Lawrence Berkeley National Laboratory,
	Berkeley, CA 94720, USA\\
	Current address: Department of Physics, Purdue University, 
	1396 Physics Building, West Lafayette, IN 47907, USA}

\maketitle

\begin{abstract}
It is found that the average baryon phase-space density obtained from 
the ratio of deuteron to proton yields is nearly constant over 
centrality in Au+Au collisions at the AGS. 
The finding offers an explanation for the puzzling centrality 
independence of the ratio of charged kaon total yields.
The correlation between the charged kaon ratio and the average baryon 
phase-space density is studied for central heavy-ion collisions 
of various systems over a wide range of beam energy. 
It is found that the charged kaon ratio and the average baryon 
phase-space density both increase with decreasing beam energy, 
and are strongly correlated. 
Such study may provide a new approach to search for 
medium effects on the kaon mass.
\end{abstract}

\pacs{PACS number(s): 25.75.-q, 25.75.Dw}


Strangeness production has been extensively studied in heavy-ion 
collisions because enhanced strangeness production may signal the 
formation of Quark-Gluon Plasma.
It has been observed that kaon production per participant nucleon 
increases with centrality in Si+Al, Si+Au, and Au+Au collisions 
at the Alternating Gradient Synchrotron (AGS),
reaching a factor of 3--4 enhancement in central collisions 
with respect to p+p interactions~\cite{E866K}. 
However, the ratio of charged kaon total yields ($\K$) 
varies little with the collision centrality~\cite{E866K}.
Similar results have been also observed in Ni+Ni collisions at 
the Schwer Ionen Synchrotron (SIS)~\cite{KaoS_NiNi} and 
Pb+Pb collisions at the Super Proton Synchrotron (SPS)~\cite{Sikler}.
The nearly constant $\K$ is puzzling because $K^+$ and $K^-$ are 
thought to be produced by different mechanisms: $K^-$'s are produced 
by pair production together with a $K^+$, while $K^+$'s can be produced, 
in addition, by associate production together with a hyperon. 
These different production mechanisms lead to different rapidity 
distributions which are observed, namely,
the $K^+$ rapidity distribution is broader than $K^-$'s~\cite{E866K}. 
One naively expects that in heavy-ion collisions the relative 
contribution of associate over pair production increases with centrality, 
because the associate production threshold for kaons is lower than 
the pair production threshold, and there are more particle re-interactions 
(at lower than the full energy) in central than peripheral collisions. 
Therefore, one expects an increasing $\K$ with centrality.

On the other hand, the constituent quark model~\cite{Anisovich,Bjorken} 
has been successfully applied to describe particle ratios in heavy-ion 
collisions~\cite{Bialas,Zimanyi}.
In the constituent quark model, 
$K^+ = u\overline{s}$ and $K^- = \overline{u}s$. 
Hence, $\K$ depends on the baryon (baryon$-$antibaryon) phase-space 
density established in the collision zone at chemical freeze-out 
(when particles cease to interact inelastically hence particle 
abundances and ratios are fixed)~\cite{Wang_APS,Wang_Kpi}.
In this picture, 
the observed constant $\K$ implies a constant baryon phase-space 
density over the collision centrality.
The extraction of the baryon phase-space density at chemical 
freeze-out heavily relies on chemical equilibrium models; 
such model studies have been so far limited to central collisions. 
On the other hand, the baryon phase-space density at final kinetic 
freeze-out (when particles cease to interact both inelastically and 
elastically) can be readily extracted from deuteron coalescence data. 
This is because deuterons, due to its small binding energy, 
are mostly formed by a proton and a neutron overlapping in phase-space, 
and cannot subsequently survive collisions with other particles.
In this Letter, we study the kinetic freeze-out average baryon 
phase-space density ($\fBth$) as a function of centrality, 
especially for Au+Au collisions at the AGS. 
We show that the $\fBth$ value in Au+Au collisions is nearly constant 
over centrality. The result may indicate that  the average baryon
phase-space densities at kinetic and chemical freeze-out ($\fBch$) 
are connected in a way that is independent of centrality. 
We further study the correlation between $\K$ and $\fBth$ 
in central heavy-ion collisions and demonstrate that $\K$ 
increases with $\fBth$ from SPS, to AGS, to SIS (decreasing beam energy).

However, there is a physics origin that may result in an 
opposite behavior, a decreasing $\K$ with increasing $\fBch$, 
and that is mass modification in nuclear medium. 
It is predicted that the $K^-$ effective mass is lower 
in nuclear medium than in free space, 
and the $K^+$ effective mass slightly higher~\cite{Brown1,Brown2}. 
A decreasing $K^-$ (and/or an increasing $K^+$) effective mass would 
yield a reduced $\K$ ratio.
In fact, the KaoS Collaboration has studied $\K$ in Ni+Ni collisions 
at the SIS and inferred a significant drop in the $K^-$ 
effective mass~\cite{KaoS_NiNi}. 
On the theory side, however, different conclusions can be reached 
depending on model assumptions~\cite{Brown3,Koch}.
In this Letter, we take a different approach, i.e., studying $\K$
as a function of $\fBth$. 
If the mass modification is large, one may observe a spectacular 
phenomenon: $\K$ increases and then decreases with $\fBth$.
Such phenomenon is not observed, however, we argue that 
this new approach may provide a more direct way to search for 
kaon medium modifications. 

The Letter is organized as follows.
First we present the rapidity distributions of the average nucleon 
phase-space density at the AGS for collisions of various centralities.
Then we extract $\fBth$ for each centrality and show that 
the value of $\fBth$ is nearly constant. 
Finally we demonstrate that $\K$ is strongly correlated with
$\fBth$ using the central collision data and discuss implications of 
the results on medium effects to the kaon mass.


In the coalescence model, the neutron phase-space density 
is related to the ratio of deuteron to proton yields (d/p) 
\cite{Siemens,Csernai,Wang_baryon}.
Following Ref.~\cite{Wang_baryon}, but not assuming identical 
neutron and proton yields, we obtain the spatial average nucleon 
phase-space density at kinetic freeze-out as
\begin{equation}
\fNy
= \frac{1+\pn}{2} \frac{1}{3} \frac{(dN/dy)_{\rm d}}{(dN/dy)_{\rm p}},
\label{Eq:fNy}
\end{equation}
where $\pn = (dN/dy)_{\rm p}/(dN/dy)_{\rm n}$, 
and $(dN/dy)_{\rm p,n,d}$ are the rapidity densities of proton, 
neutron, and deuteron, respectively. 
The first term arises from isospin asymmetry and will be referred to as
\begin{equation}
\alpha_I = (1+\pn)/2.
\label{Eq:alphaI}
\end{equation}
The value of $\pn$ depends on the combination of the projectile 
and target species, the collision centrality and nucleon rapidity.
The value of $\pn$ may vary little with nucleon transverse momentum
because of the isospin symmetry of strong interaction.

The deuteron and proton data from 
Si+Al and Si+Au at 14.6 AGeV/$c$~\cite{E802d} 
and Au+Au at 11.6 AGeV/$c$~\cite{E866d} are used to extract $\fNy$.
We assume $\pn = 1$ for the nearly isospin symmetric Si+Al collisions.
For Si+Au and Au+Au, the values of $\pn$ are estimated using the 
Relativistic Quantum Molecular Dynamics ({\sc rqmd}) model~\cite{Sorge}.
For the rapidity ranges covered by the data, the model indicates
$0.8< \pn < 1$, hence $0.9 < \alpha_I < 1$ for both systems.

\begin{figure}[hbt]
\centerline{
\epsfxsize=0.45\textwidth\epsfbox[10 50 510 480]{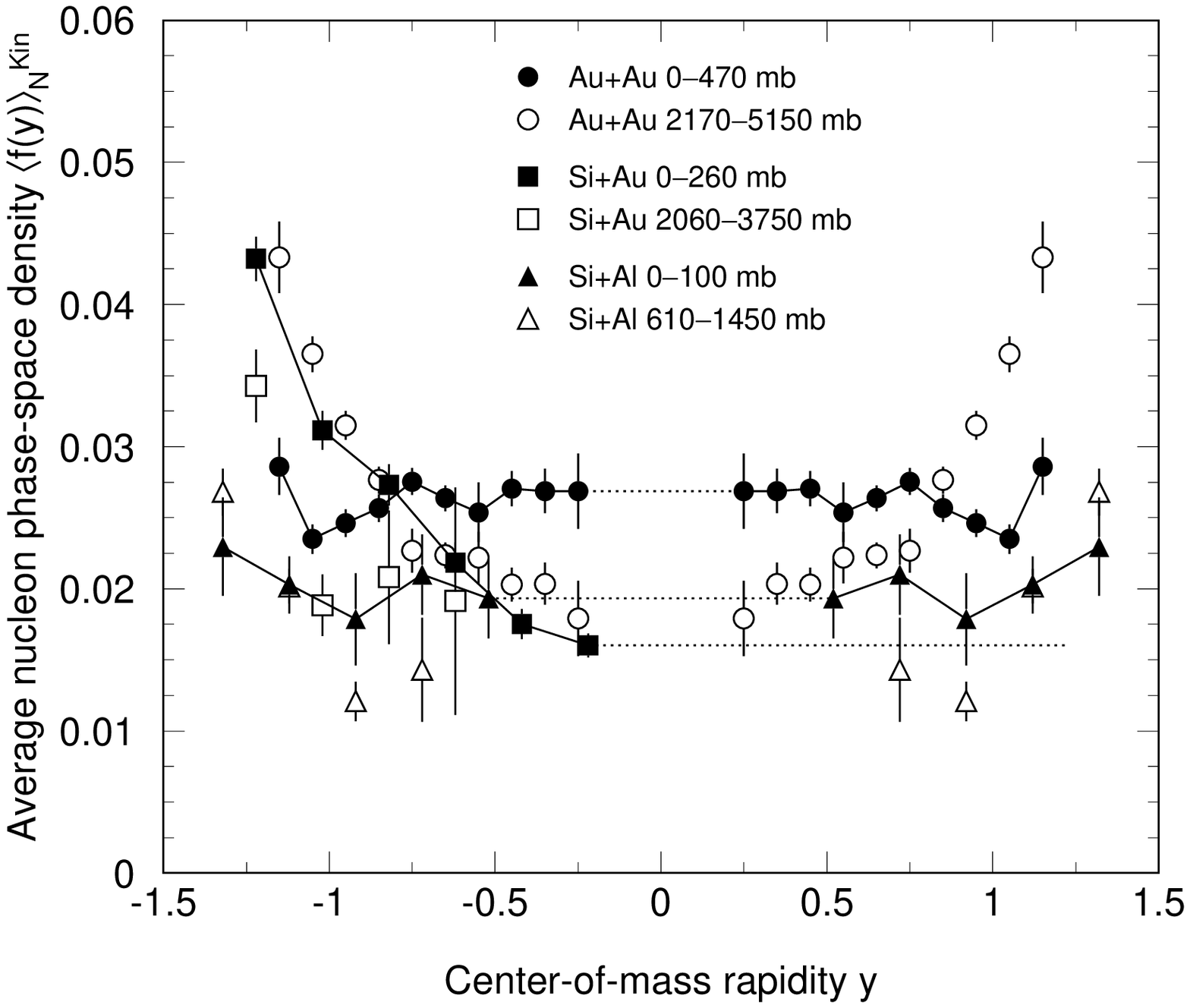}}
\caption{The spatial average nucleon phase-space density at kinetic freeze-out,
	$\fNy$, as a function of the proton center-of-mass rapidity 
	($y$) in central and peripheral heavy-ion collisions at the AGS.
	Collision centralities are indicated by the cross section ranges.
	The data points at backward rapidities ($y<0$) are measured, 
	and are reflected about mid-rapidity ($y=0$).
	Errors shown are statistical only. 
	The experimental systematic effects largely cancel in the d/p ratio.
	The solid lines connecting the data points are to guide the eye.
	The dotted lines are an extrapolation to the unmeasured regions.}
\label{fig1}
\end{figure}

Figure~\ref{fig1} shows the obtained $\fNy$ as a function 
of the proton rapidity.
$\fNy$ varies little with rapidity in central Si+Al and Au+Au collisions. 
In peripheral Si+Al and Au+Au and both central and peripheral Si+Au 
collisions, $\fNy$ is larger at target rapidity than mid-rapidity.
We note that at target rapidity deuterons may be produced by
mechanisms other than coalescence
(e.g. target fragmentation)~\cite{E866d} , 
resulting in an overestimate of $\fNy$ by Eq.~(\ref{Eq:fNy}).
The magnitude of this possible effect 
is not investigated in the present work.

The average baryon phase-space density is the relevant quantity. 
Hyperons are the most significant contributors to the baryons 
besides nucleons.
The abundance of antibaryons is negligible at AGS energies, 
and is small at SPS energies.
Generally, the average baryon phase-space density is obtained from 
\begin{equation}
\fBth = \fN (1-\overline{N}/N) [1+(Y-\overline{Y})/(N-\overline{N})],
\label{Eq:fB0} 
\end{equation}
where $\fN$ is the rapidity averaged nucleon phase-space density
(see below), and $N, \overline{N}, Y$, and $\overline{Y}$ are the nucleon,
antinucleon, hyperon, and antihyperon total yields, respectively. 
It has been assumed that the phase-space 
distributions of hyperons and antibaryons are the same as of 
nucleons~\cite{dbar}. 
This assumption does not introduce significant errors on $\fBth$ 
as the contributions of hyperons and antibaryons are generally small 
(see Table~\ref{tab}).

We denote
\begin{equation}
\alpha_N = 1-\overline{N}/N \approx 1-\overline{\rm p}/{\rm p}
\label{Eq:alphaN} 
\end{equation}
where $\overline{N}/N$ has been approximated by the
antiproton to proton ratio ($\overline{\rm p}/{\rm p}$),
and
\begin{equation}
\alpha_Y = 1+(Y-\overline{Y})/(N-\overline{N}). 
\label{Eq:alphaY}
\end{equation}
Hence, 
\begin{equation}
\fBth = \fN \cdot \alpha_N \cdot \alpha_Y.
\label{Eq:fB}
\end{equation}
The conservations of global baryon number and strangeness give, 
respectively,
\begin{equation}
N-\overline{N} = \npa - (Y-\overline{Y}) 
\label{Eq:Blaw}
\end{equation}
where$\npa$ is the total number of participant nucleons, and
\begin{equation}
Y-\overline{Y} \approx K-\overline{K} = (1+\alpha_K)\cdot(K^+ - K^-) 
\label{Eq:Slaw}
\end{equation}
where $\alpha_K = (K^0 - \overline{K^0})/(K^+ - K^-)$.
Therefore, Eq.~(\ref{Eq:alphaY}) becomes
\begin{equation}
\alpha_Y = [1-(1+\alpha_K)\cdot(K^+ - K^-)/\npa]^{-1}.
\label{Eq:alphaY2}
\end{equation}
The value of $\alpha_K$ is not one in isospin asymmetric collisions, 
however, we will use $\alpha_K \approx 1$ because the effect 
of a non-unitary value~\cite{alpha_K} 
is reduced by $(K^+ - K^-)/\npa$ which is a small quantity 
for collisions we consider. 
It should be noted that 
only one unit of strangeness per multi-strange hyperon is counted
in Eq.~(\ref{Eq:Slaw}).
This is safe because the production of multi-strange hyperons
is relatively small even at the SPS energies~\cite{NA49Xi}.

The AGS E859/E866 proton and deuteron measurements cover 
a broad rapidity range exploiting 
the (near) symmetry of the Si+Al and Au+Au systems~\cite{broad_y}. 
We take the $(dN/dy)_{\rm p}$ weighted average of $\fNy$ 
in Eq.~(\ref{Eq:fNy}) over this rapidity range to obtain $\fN$.
The unmeasured d/p ratio in the mid-rapidity region is approximated 
by the dotted lines in Fig.~\ref{fig1} connecting the last point
(i.e., closest to mid-rapidity) with the reflected one.
For Si+Au we assume that the unmeasured d/p ratio 
at the more forward rapidities is the same as the last data point.
We assign a 8\% systematic error on $\fN$ for Si+Au 
due to this extrapolation by examining a smoothly dropping d/p ratio
from the low rapidity data points.
From $\fN$, we extract $\fBth$
using Eqs.~(\ref{Eq:fB}), (\ref{Eq:alphaN}), and (\ref{Eq:alphaY2}). 
For all three systems $\alpha_N \approx 1$ and $\alpha_Y$ ranges 
from 1.03 in peripheral to 1.13 in central collisions.
For Si+Al and Au+Au, we assign a 5\% systematic error on $\fBth$
due to the assumptions made to extract $\fBth$ and errors on the
correction factors. For Si+Au, we assign a 10\% systematic error on
$\fBth$.

\begin{figure}[hbt]
\centerline{
\epsfxsize=0.5\textwidth\epsfbox[10 40 565 480]{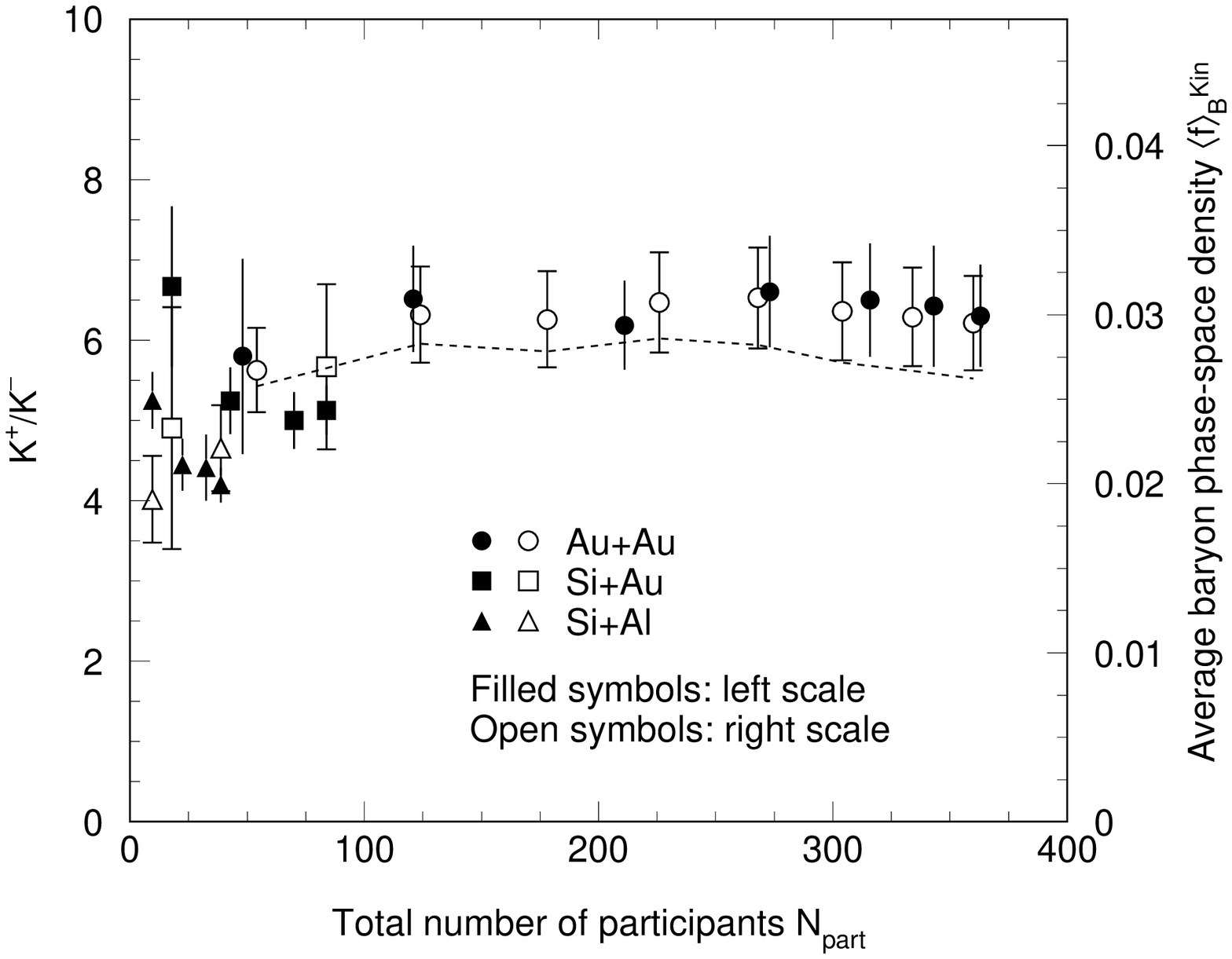}}
\caption{The $\K$ ratio (filled symbols, left scale) and the average 
	baryon phase-space density at kinetic freeze-out $\fBth$ 
	(open symbols, right scale) as a function of the 
	total number of participant nucleons ($\npa$) at the AGS.
	Errors shown are statistical for $\K$ and statistical and
	systematic errors added in quadrature for $\fBth$.
	The experimental systematic effects largely
	cancel in the $\K$ and d/p ratios.
	Note that both quantities are nearly constant over centrality.
	The dashed curve shows the average nucleon phase-space density 
	at kinetic freeze-out (right scale); 
	the errors are comparable to the $\fBth$ ones.}
\label{fig2}
\end{figure}

Figure~\ref{fig2} shows $\K$~\cite{E866K} and $\fBth$ 
at the AGS as a function of $\npa$. 
Both quantities are nearly constant.
It should be noted that the constant values do not necessarily 
demonstrate that $\K$ and $\fBth$ are correlated. 
However, it will be shown that they are correlated 
in central collisions at different energies. 
Therefore, it is fair to argue that they are correlated also in the 
same collision system and are independent of the collision centrality. 
Because $\K$ is fixed at chemical freeze-out, the results may imply that 
the $\fBth$ values at kinetic freeze-out, which presumably happens later 
than chemical freeze-out in heavy-ion collisions, are connected to the 
chemical freeze-out $\fBch$ values in a way that is independent of centrality.

Since the difference between the kaon yields ($K^+ - K^-$) enters 
into $\fBth$ as a correction, one may be concerned about auto-correlation.
However, such concern is unnecessary because 
(1) no correlation is present between $\K$ and $K^+ - K^-$ 
(which increases with $\npa$ more than linearly~\cite{E866K}),
and (2) the $K^+ - K^-$ correction to $\fBth$ is small (at most 13\%).
In fact, one may correlate $\K$ with the average nucleon phase-space 
density which is shown as the dashed curve in Fig.~\ref{fig2}.

It should be noted that the correlation exists only 
between the rapidity integrated values of $\fBth$ and $\K$; 
locally in rapidity, $\fBth(y)$ and $\K(y)$ are not or much less correlated 
(especially in central Au+Au collisions).
This may be due to longitudinal flow which can be different for $K^+$,
$K^-$, and nucleon~\cite{Braun}. 
Similar problems are encountered in equilibrium model
studies of particle ratios: thermodynamic relations between various 
particle ratios are only valid for ratios of integrated yields; 
differentially they do not hold because of longitudinal and 
transverse flow~\cite{Cleymans2}.

Now we study the correlation between $\K$ and $\fBth$ 
in central collisions at different beam energies.
The values of $\K$ and $\fBth$ extracted from experimental data 
are tabulated in Table~\ref{tab}.
The $\K$ values for S+S~\cite{NA44K} and Ni+Ni~\cite{KaoS_NiNi} 
are at mid-rapidity. 
The $\fBth$ values for S+S~\cite{Murray} and Pb+Pb~\cite{Murray,Hansen} 
are determined from d/p at mid-rapidity. 
These values are approximated as for the total yield
ratios with the systematic errors quoted in Table~\ref{tab}. 
The $\K$ and $\fBth$ values for Au+Au~\cite{E866K,E866d} and
Ni+Ni~\cite{KaoS_NiNi,FOPI} are obtained from measurements 
at slightly different beam energies. 
For Ni+Ni we have assumed $\pn = 1$. 

\begin{figure}[hbt]
\centerline{
\epsfxsize=0.45\textwidth\epsfbox[10 50 510 480]{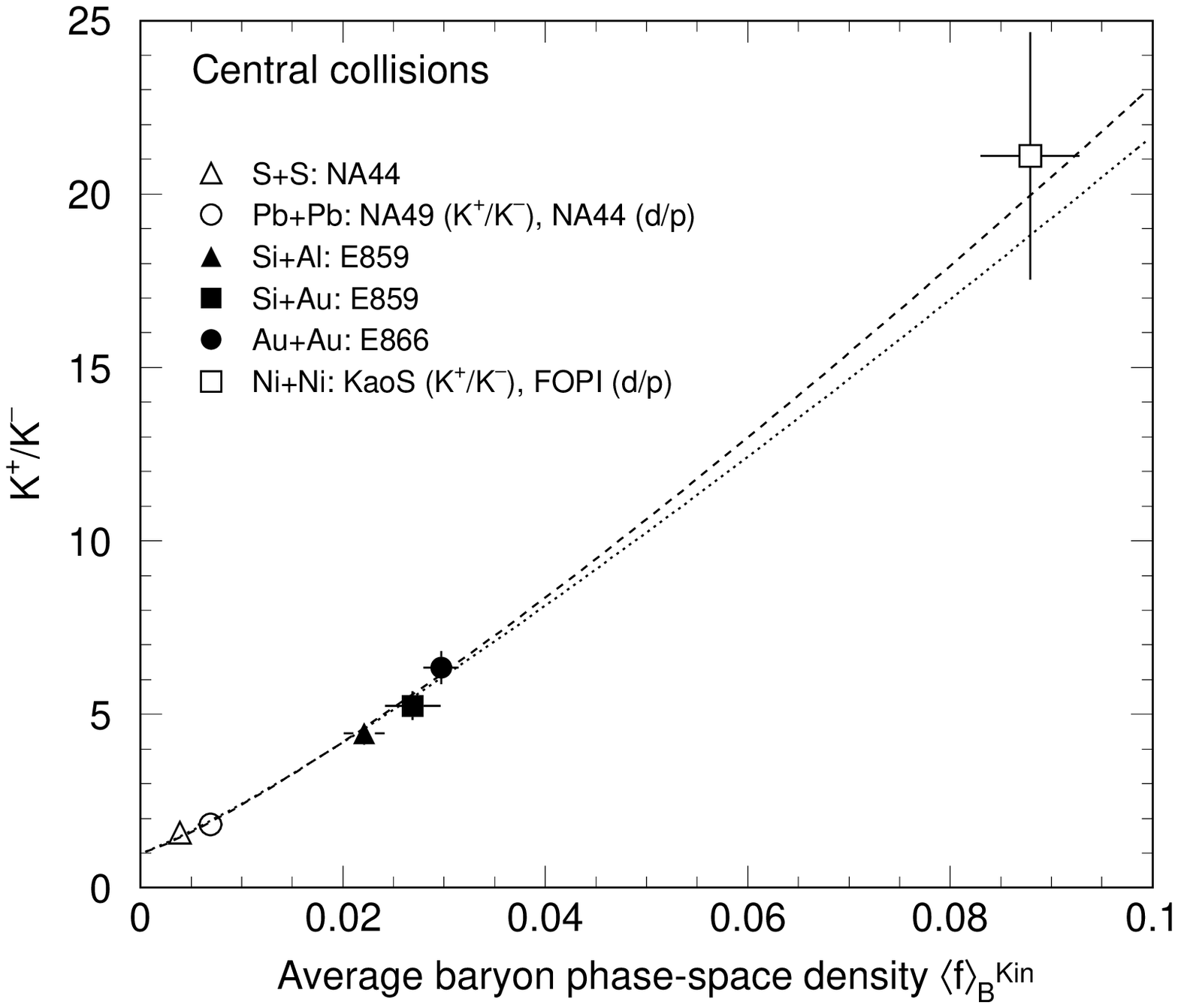}}
\caption{The $\K$ ratio as a function of the average baryon phase-space density 
	at kinetic freeze-out $\fBth$ in central heavy-ion collisions.
	Errors shown are statistical and systematic errors added in 
	quadrature. The data are tabulated in Table~I.
	The dashed curve is a fit to power-law function. 
	The dotted curve is a similar fit 
	but excluding the Ni+Ni data point.}
\label{fig3}
\end{figure}

Figure~\ref{fig3} shows $\K$ as a function of $\fBth$.
Both quantities decrease with beam energy. 
The data from various collision systems at different beam energies 
follow a similar dependence.
Motivated by equilibrium models, we fit the data to power-law function, 
yielding
\begin{equation}
\frac{K^+}{K^-} = 1 + 
\left( \frac{\fBth}{0.0076 \pm 0.0009} \right)^{1.20 \pm 0.10}. 
\label{Eq:power}
\end{equation}
The fit result is plotted in Fig.~\ref{fig3} as the dashed curve.

The KaoS Collaboration observed a $\K$ ratio in Ni+Ni collisions 
at the same equivalent (sub-threshold) beam energy that is about 
seven times smaller than in p+p near threshold. 
The KaoS Collaboration contributed this difference to a drop
in the $K^-$ mass in Ni+Ni collisions, and inferred a mass drop of 
$270^{+55}_{-90}$~MeV/$c^2$ by using the measured kaon excitation 
function~\cite{KaoS_NiNi}. 
Since $\fBth$ is a measure of matter density, we argue that 
studying $\K$ against $\fBth$ may provide a different and probably 
more direct approach to search for medium effect on kaon mass: 
If the $K^-$ mass is modified by medium at low energy
but not (or differently) at high energy, 
then the low energy $\K$ should 
deviate from extrapolation of the high energy data.
This Letter is a first attempt of this new approach:
The fit result in Fig.~\ref{fig3} seems to imply no difference 
between the SIS and AGS/SPS data; fitting the data
excluding the Ni+Ni point to power-law function, yielding the result
shown in the dotted curve, seems to give the same implication.

Moreover, as noted, the average baryon phase-space densities at chemical 
and kinetic freeze-out may be different; 
the more suitable variable to correlate with $\K$ 
is the chemical freeze-out one.
It has been found in Ref.~\cite{Wang_baryon}, under the assumption of 
chemical and thermal equilibrium, that the chemical freeze-out $\fBch$ 
is twice the kinetic freeze-out $\fBth$ at both the AGS and SPS.
Using the thermal model parameters obtained from the Ni+Ni 
data~\cite{Cleymans}, the $\fBch$ value is found 
to be the same as the $\fBth$ value.
In Fig.~\ref{fig3}, if the $\fBch$ values were plotted, 
then the Ni+Ni data point would lie above a sensible extrapolation from 
the AGS and SPS data by about a factor of two. 
This is even more contradictory to $K^-$ mass drop, 
as $K^-$ mass drop would give a reduced $\K$ ratio.

However, one should be careful to conclude from the above results
that there is no $K^-$ mass drop in the Ni+Ni data. 
This is because: 
\begin{enumerate}
\item 
Although the power law parameterization is motivated
by equilibrium models, it is not a priori for the relation
between $\K$ and $\fBth$.
In fact, data from C+C collisions at the SIS~\cite{KaoS_CC} 
do not follow the parameterization shown in Fig.~\ref{fig3}: 
there a large $\K$ ratio of $39\pm 6$ was observed; 
no d/p data is available, but it can be conjectured 
that the $\fBth$ value is lower in C+C than in Ni+Ni.
It is possible that the C+C data follow a different systematic 
consistent with the AGS/SPS data, while the Ni+Ni data lie 
below this systematic. 
This would be an interesting result because 
it implies a $K^-$ mass drop in Ni+Ni collisions.
\item
Since the leverage of the AGS and SPS data is small, the systematic
errors on these data are critically important to the extrapolation to the
SIS data. 
\end{enumerate}
In order to address these concerns, more data are needed to fill
in the gap between AGS and SIS.
These data are underway from the E895 Collaboration~\cite{Rai}. 
New data from the SPS at 40 AGeV would be also valuable 
as they fill in the gap between the full-energy AGS and SPS data.

In summary, we have extracted the average baryon phase-space density 
at kinetic freeze-out from the deuteron to proton ratio in Si+Al, 
Si+Au and Au+Au collisions at the AGS~\cite{E802d,E866d}. 
The average baryon phase-space density at kinetic freeze-out
is found to be nearly constant over the collision centrality, 
which offers an explanation for the puzzling centrality independence 
of the charged kaon total yield ratio~\cite{E866K}.
The correlation between the charged kaon ratio and the average baryon 
phase-space density is studied for central heavy-ion collisions of 
various systems over a wide range of beam energy. 
It is found that the charged kaon ratio increases 
strongly with the average baryon phase-space density, and
both variables increase with decreasing beam energy.
It is argued that studying the correlation between these variables 
may provide a different and 
more direct approach to search for medium effect on kaon mass.
Although no conclusion can be drawn from the present study regarding 
medium modification to the kaon mass in Ni+Ni collisions 
at the SIS~\cite{KaoS_NiNi}, new data from the AGS low energy collisions 
should shed light on this question.


The author gratefully acknowledges discussions with Drs. 
C. Chasman, B.A. Cole, E. Garcia, F. Laue, G. Odyniec, G. Rai, 
H.G. Ritter, M.J. Tannenbaum, and N. Xu. 
He also thanks G. Cooper for the critical reading of the manuscript.
This work was supported by the U.S. Department of Energy 
under contract DE-AC03-76SF00098.


\newpage

\begin{minipage}{\textwidth}

\begin{table}
\caption{The $\K$ ratio and the average baryon phase-space density at
	kinetic freeze-out $\fBth$ (extracted from d/p ratios) in central 
	heavy-ion collisions. The factors $\alpha_I$, $\alpha_N$, and $\alpha_Y$ 
	take into account, respectively, the effects of isospin asymmetry, 
	antinucleon and (anti)hyperon yields.
	For $\K$ and $\fBth$ the first error is statistical. 
	The systematic errors (in percentage) are introduced by the 
	extrapolation in rapidity, the uncertainties in 
	$\alpha_I$, $\alpha_N$, and $\alpha_Y$,
	and the preliminary nature of some of the data. 
	The experimental systematic effects largely cancel
	in the $\K$ and d/p ratios.}
\label{tab}

\begin{tabular}{cll|ll|lllll}
Collision & Beam Energy	& Centrality & $\K$ & 
			& $\alpha_I$ & $\alpha_N$ & $\alpha_Y$ & $\fBth$ &\\
System    & per Nucleon	& (mb)       & & & & & &\\ 
\hline 
S + S & 200 GeV & 160	& $1.58   \pm 0.09   \pm  5\%$ 	& \cite{NA44K}
			& 1	& 0.89 \cite{NA44pbar}	& 1.32$^b$ 
			& $0.0039 \pm 0.0007 \pm 10\%$	& \cite{Murray}\\
Pb+Pb & 158 GeV & 340	& $1.83   \pm 0.02   \pm  5\%$	& \cite{Sikler}
			& 0.97 \cite{Sorge}	& 0.93 \cite{Sikler}	& 1.32	
			& $0.0069 \pm 0.0010 \pm 10\%$	& \cite{Murray,Hansen}\\
Si+Al &14.6 GeV/$c$ &100& $4.45   \pm 0.32           $	& \cite{E866K}
			& 1	& 1	& 1.11
			& $0.0221 \pm 0.0017 \pm 5\% $	& \cite{E802d}\\
Si+Au &14.6 GeV/$c$&260	& $5.25   \pm 0.42           $	& \cite{E866K}
			& 0.95 \cite{Sorge}	& 1	& 1.13	
			& $0.0269 \pm 0.0006 \pm 10\%$	& \cite{E802d}\\
Au+Au &11.1, 11.6$^a$ GeV/$c$	& 420, 470$^a$ 
			& $6.35 \pm 0.48$ & \cite{E866K}
			& 0.96 \cite{Sorge}	& 1	& 1.12	
			& $0.0297 \pm 0.0009 \pm 5\% $	& \cite{E866d}\\
Ni+Ni & 1.8, 1.93$^a$ GeV & 100 
			& $21.1   \pm 3.4    \pm  5\%$	& \cite{KaoS_NiNi}
			& 1	& 1	& 1	
			& $0.0879 \pm 0.0021 \pm  5\%$	& \cite{FOPI} 
\end{tabular}
$^a$ 
The first number is for the $\K$ data; the second is for the d/p data. 
Quoted for Ni+Ni are the kinetic beam energies. 
$^b$ 
The Pb+Pb $\alpha_Y$ value is used because 
the total yield $K^+ - K^-$ is not available for S+S.
\end{table}

\end{minipage}


\begin{thebibliography}{10}

\bibitem{E866K}
E802 Collaboration, L. Ahle {\it et al.}, 
Phys. Rev. C {\bf 60} (1999) 044904.

\bibitem{KaoS_NiNi}
KaoS Collaboration, R. Barth {\it et~al.}, 
Phys. Rev. Lett. {\bf 78} (1997) 4007.

\bibitem{Sikler}
F. Sikler (NA49 Collaboration), Nucl. Phys. {\bf A661} (1999) 45c.

\bibitem{Anisovich}
V.V. Anisovich and V.M. Shekhter, Nucl. Phys. {\bf B55} (1973) 455;
{\it ibid} {\bf B63} (1973) 542(E).

\bibitem{Bjorken}
J.D. Bjorken and G.R. Farrar, Phys. Rev. D {\bf 9} (1974) 1449.

\bibitem{Bialas}
A. Bialas, Phys. Lett. B {\bf 442} (1998) 449.

\bibitem{Zimanyi}
J. Zim\'anyi {\it et al.}, Phys. Lett. B {\bf 472} (2000) 243.

\bibitem{Wang_APS}
F. Wang, in proceedings of the Relativistic Heavy Ion Mini-Symposium 
at the APS Centennial, Atlanta, 1999 [nucl-ex/9905005].

\bibitem{Wang_Kpi}
F. Wang {\it et~al.}, Phys. Rev. C {\bf 61}, 064904 (2000).

\bibitem{Brown1}
G.E. Brown and M. Rho, Phys. Rev. Lett. {\bf 66} (1991) 2720.

\bibitem{Brown2}
G.E. Brown {\it et al.}, Phys. Rev. C {\bf 43} (1991) 1881.

\bibitem{Brown3}
G.Q. Li and G.E. Brown, Phys. Rev. C {\bf 58}, 1698 (1998).

\bibitem{Koch}
J. Schaffner-Bielich, V. Koch, and M. Effenberger, 
Nucl. Phys. {\bf A669}, 153 (2000). 

\bibitem{Siemens}
P.J. Siemens and J.I. Kapusta, Phys. Rev. Lett. {\bf 43} (1979) 1486;
{\it ibid} {\bf 43} (1979) 1690(E).

\bibitem{Csernai}
L.P. Csernai and J.I. Kapusta, Phys. Rep. {\bf 131} (1986) 223.

\bibitem{Wang_baryon}
F. Wang and N. Xu, Phys. Rev. C {\bf 61} (2000) 021904(R).

\bibitem{E802d}
E802 Collaboration, T. Abbott {\it et al.}, 
Phys. Rev. C {\bf 50} (1994) 1024.

\bibitem{E866d}
E802 Collaboration, L. Ahle {\it et al.}, 
Phys. Rev. C {\bf 60} (1999) 064901.

\bibitem{Sorge}
H. Sorge, Phys. Rev. C {\bf 52} (1995) 3291.

\bibitem{dbar}
Without this assumption, one has to use 
${\rm d}/{\rm p}-\overline{\rm d}/\overline{\rm p}$ 
to deduce the nucleon phase-space density. 
This becomes difficult as $\overline{\rm d}$ is involved 
whose experimental measurements are poor.

\bibitem{alpha_K}
The {\sc rqmd} model indicates that $1<\alpha_K<1.05$ 
in central Au+Au or Pb+Pb collisions at the AGS and SPS. 

\bibitem{NA49Xi}
NA49 Collaboration, H.~Appelsh\"{a}user {\it et~al.}, 
Phys. Lett. B {\bf 444} (1998) 523.

\bibitem{broad_y}
The maximum rapidity of a nucleon in a nucleus due to Fermi motion 
is about 0.26. The projectile Si and Au rapidities at the AGS are
about 1.72 and 1.6, respectively. Thus the measurements cover
almost the whole rapidity range beyond Fermi motion.

\bibitem{Braun}
P.~Braun-Munzinger {\it et~al.}, Phys. Lett. B {\bf 344}, 43 (1995).

\bibitem{Cleymans2}
J. Cleymans, H. Oeschler, and K. Redlich, J. Phys. G {\bf 25}, 281 (1999).

\bibitem{NA44K}
NA44 Collaboration, H.~B$\o$ggild {\it et~al.}, 
Phys. Rev. C {\bf 59} (1999) 328.

\bibitem{NA44pbar}
NA44 Collaboration, I.~G.~Bearden {\it et~al.}, 
Phys. Rev. C {\bf 57} (1998) 837.

\bibitem{Murray}
M. Murray (NA44 Collaboration), Nucl. Phys. {\bf A661} (1999) 456c.

\bibitem{Hansen}
A.G. Hansen (NA44 Collaboration), Nucl. Phys. {\bf A661} (1999) 387c.

\bibitem{FOPI}
FOPI Collaboration, B. Hong {\it et~al.}, 
Phys. Rev. C {\bf 57} (1998) 244.

\bibitem{Cleymans}
J. Cleymans {\it et al.}, Phys. Rev. C {\bf 57} (1998) 3319.

\bibitem{KaoS_CC}
KaoS Collaboration, F. Laue {\it et~al.}, 
Phys. Rev. Lett. {\bf 82}, 1640 (1999).

\bibitem{Rai}
G. Rai and C. Ogilvie, private communications.

\end{thebibliography}
\end{document}